\documentclass[prl,twocolumn,showpacs,amsmath,superscriptaddress,runinaddress,10pt,letterpaper,floatfix]{revtex4}

\usepackage{upgreek} 
\usepackage{graphicx}
\usepackage{subfigure}
\usepackage{times}
\usepackage{txfonts}
\usepackage{amssymb}
\usepackage{booktabs}

\newcommand{\SD}{$\mathrm{SD}_\mathrm{geo}$}

\begin{document}

\title{On the origin and extent of  mechanical variation among cells}
\author{John M. Maloney}
\affiliation{%
Department of Materials Science and Engineering
}
\author{Krystyn J. Van Vliet$^{1,}$}%
\thanks{Address: Laboratory for Material Chemomechanics (8-237); Massachusetts Institute of Technology; 77 Massachusetts Ave.; Cambridge, MA 02139}
\email{krystyn@mit.edu}
\affiliation{%
Department of Biological Engineering \\
Massachusetts Institute of Technology,
Cambridge, MA 02139 USA\\
}

\begin{abstract}
Investigations of natural variation in cell mechanics within a cell population are essential to understand the stochastic nature of soft-network deformation. Striking commonalities have been found concerning the average values and distribution of rheological parameters of cells: first, attached and suspended cells exhibit power-law rheological behavior; second, cell stiffness is distributed log-normally. A predictive connection between these two near-universal findings has not been reported, to our knowledge.  Here we postulate, based on our own and others' experimental reports and leading models of cell rheology, that the exponent that characterizes power-law rheology varies intrinsically among cells as an approximately Gaussian-distributed variable. Besides explaining naturally the log-normal distribution of cell stiffness that is widely observed, this postulate predicts multiple empirically observed relationships from cell deformation studies. Our framework ultimately links inherent noise in postulated relaxation mechanisms of cytoskeletal networks to mechanical variation among cells and  cell populations.\end{abstract}

\pacs{87.17.Rt, 87.16.Ln, 83.80.Lz, 05.40.-a}
\maketitle

\begin{figure}[b!]
\centering
\includegraphics[width=8.7cm]{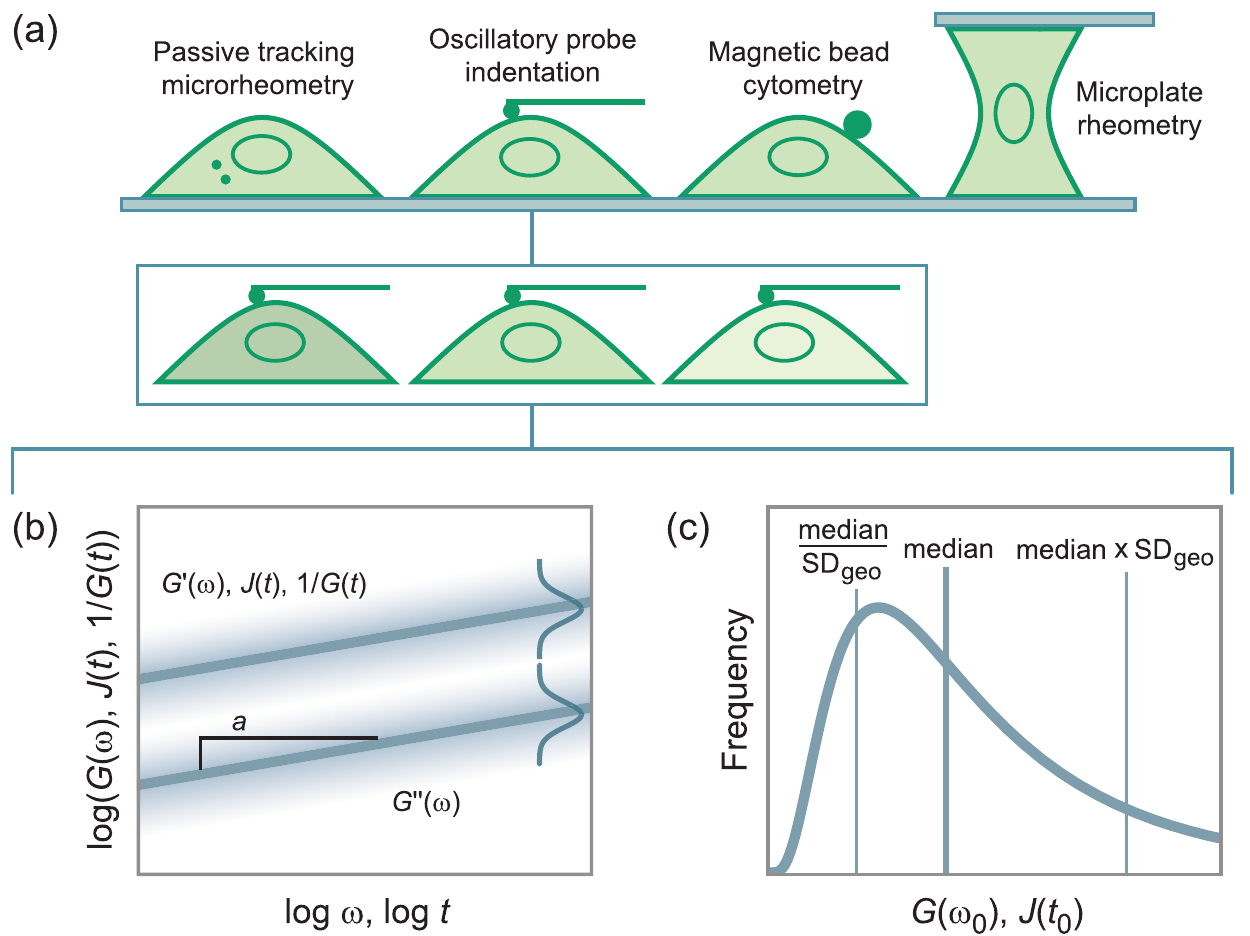}
\caption{(a) Universal mechanical behavior of
living cells at multiple length scales: (b) The storage and
loss moduli $G^\prime(\omega)$ and $G^{\prime\prime}(\omega)$, along with the creep compliance $J(t)$ and
reciprocal stress relaxation modulus $1/G(t)$, scale as a power law with
frequency $\omega$ or time $t$, respectively; (c)  Individual measurements of
these rheological parameters are distributed log-normally. 
}\label{fig:PLR-SDgeo1}
\end{figure}

In light of the complexity and structural heterogeneity of biological cells, findings of universal mechanical tendencies are significant. Thus, it is intriguing that, on physiologically relevant time scales (at least three orders of magnitude centered on 1\,s), and in experiments at multiple length scales (Fig.~\ref{fig:PLR-SDgeo1}(a)),  animal cells regularly exhibit so-called power-law rheology (Fig.~\ref{fig:PLR-SDgeo1}(b)). For example, dynamic cell stiffness (storage and loss moduli $G^\prime(\omega)$ and $G^{\prime\prime}(\omega)$) scales with frequency $\omega$ as $G^\prime(\omega),G^{\prime\prime}(\omega)\propto \omega^{a}$~\cite{fabry2001scaling,lenormand2004linearity,puig-cytoskeletal,massiera2007mechanics}, and creep compliance $J(t)$~\cite{lenormand2004linearity,balland2006power,kollmannsberger-nonlinear,maloney2010} and stress relaxation modulus $G(t)$~\cite{hemmer2009role} scale with time $t$ as $J(t),1/G(t)\propto t^{a}$, with $a\approx 0.1$--0.3 typically. This behavior is attributed to the varied possible arrangements in cytoskeletal networks, each with its own relaxation time, that integrate to form a material with no single characteristic time scale~\cite{fredberg2006cytoskeleton}.

Common findings also exist regarding the distribution of individual cell stiffness values around a population average. Cell stiffness is distributed log-normally---again largely independent of experimental technique and length scale~\cite{fabry2001scaling,desprat2005creep,hoffman2006consensus,hiratsuka2009number} and also independent of metabolic state~\cite{hoffman2006consensus} and cytoskeletal perturbation~\cite{fabry2003time,puig-cytoskeletal,lenormand2004linearity} (Fig.~\ref{fig:PLR-SDgeo1}(c)). Wide population distributions of stiffness measurements have previously been attributed to probe contact variation~\cite{fabry2001scaling,balland2006power}, but arise even with  non-contact single-cell deformation~\cite{maloney2010}. In contrast to the existence of multiple models of power-law rheology (PLR, reviewed in \cite{hoffman2009cell}), the observed log-normal distribution has remained largely uninvestigated and especially lacks any predictive explanation. (A previous report replicates the distribution shape via a random selection of viscoelastic elements, but does not provide a way to predict distribution parameters~\cite{balland2006power}.)   Also unexplained are observed differences in the width of distributions (quantified by geometric standard deviation \SD) between storage and loss modulus measurements~\cite{hiratsuka2009number}. As we develop a better understanding of the cell as a mechanical material, natural questions are (1) whether PLR implies a log-normal distribution (and if so, under what necessary postulates); (2) what is the minimum measurable mechanical variation in a cell population; and (3) whether fluctuations in the energy landscapes of cytoskeletal networks can be linked to mechanical variation among cells. 

\begin{figure}
\includegraphics[width=8.7cm]{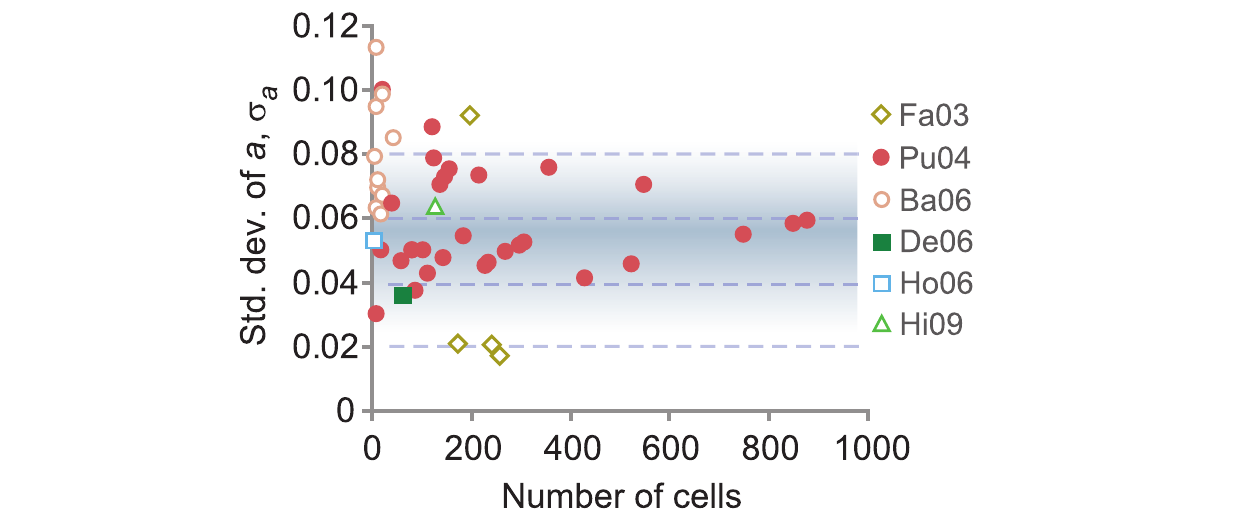}
\caption{Existing estimates of standard deviation  $\sigma_a$ of power-law exponent $a$ as a function of data set size  suggest that the inherent uncertainty in $a$ lies between 0.02--0.10. Data referenced by study author and year~\cite{fabry2003time,puig-cytoskeletal,
balland2006power,deng2006fast,hoffman2006consensus,hiratsuka2009number,kollmannsberger-nonlinear}.}\label{fig:PLR-SDgeo3}
\end{figure}

With an interest in answering these questions, we propose that the power-law exponent $a$ varies \emph{intrinsically} among cells approximately as a Gaussian-distributed variable, based on the following evidence:  First, experimental measurements of the exponent $a$ have actually been found to exhibit a Gaussian distribution~\cite{desprat2005creep,balland2006power,massiera2007mechanics,hiratsuka2009number}. Moreover,  literature reports on the variation of $a$ (quantified as standard deviation $\sigma_a$), as shown in Fig.~\ref{fig:PLR-SDgeo3}, suggest an intrinsic variation that lies in the approximate range 0.02--0.10 and is sustained even when hundreds of individual cells are sampled
. Second, multiple PLR-predicting models of mechanical networks relate $a$ to an energy. (For example, in the soft glassy rheology (SGR) model developed by Sollich et al.~\cite{sollich1997rheology} and applied to cells by Fabry and Fredberg et al.~\cite{fabry2001scaling,fredberg2006cytoskeleton}, $x=a+1$ is the ratio of the average agitation energy to the energy associated with a glass transition; in the glassy wormlike chain model, $\mathcal{E}\approx 3/a$ is the average energy barrier retarding material relaxation~\cite{semmrich2007glass}.) Notably, these models assume that the power-law exponent represents the sum of many independent energies~\cite{sollich1997rheology,semmrich2007glass}; from the central limit theorem, therefore, we would expect the exponent $a$ to be approximately Gaussian regardless of the distributions of its constitutive energy components.  Thus, our  postulate of intrinsic Gaussian exponent appears plausible in both experimental and theoretical contexts. We further assume that endogenous variation in $a$ dominates over other sources of variation such as direct engagement of the cytoskeleton via adhesion complexes~\cite{puig-cytoskeletal}.

\begin{figure}[b!]
\centering
\includegraphics[width=8.7cm]{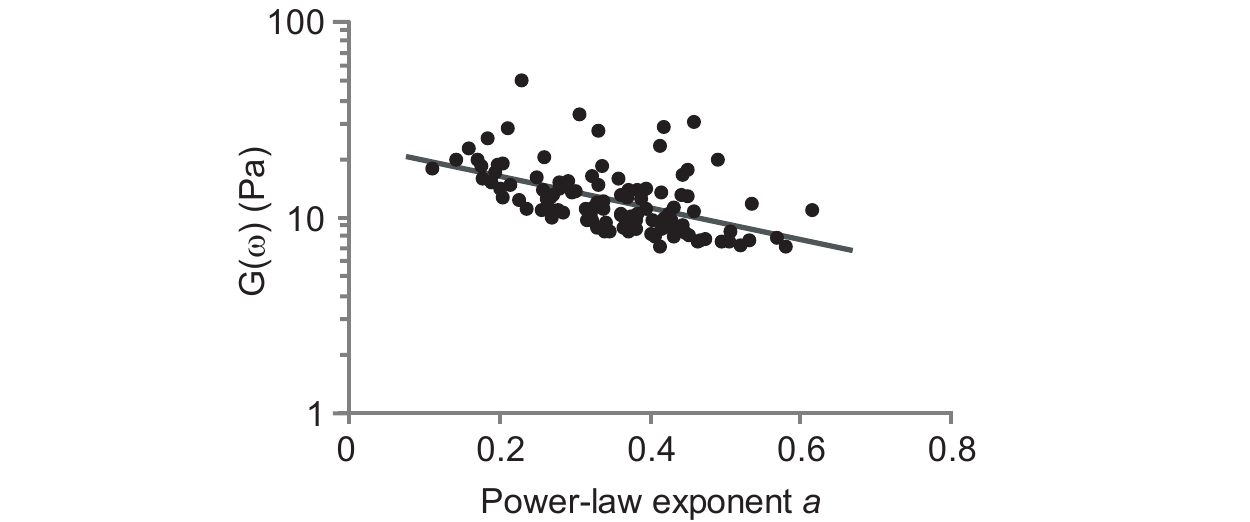}
\caption{In non-contact measurements of individual mesenchymal stem cells by optical stretching, stiffness $G$ is negatively correlated with power-law exponent $a$ for sinusoidal measurements at 0.5\,Hz.}
\label{fig:G-vs-a}
\end{figure}

We now relate postulated fluctuations in the power-law exponent $a$ to quantitative predictions of variation in mechanical parameters such as the complex modulus $G^\star(\omega)=G^\prime(\omega)+iG^{\prime\prime}(\omega)$, creep compliance $J(t)$, and stress relaxation modulus $G(t)$. Consider first the complex modulus and its storage and loss components, expressed as $f(a)(\omega/Y_0)^{a}$, where $f(a)$ is a characteristic prefactor for that component (namely, $\Gamma(1+a)\Gamma(1-a)\cos[\pi (a/2]$ for $G^\prime$ and $\Gamma(1+a)\Gamma(1-a)\sin[\pi a/2]$ for $G^{\prime\prime}$) as obtained from the structural damping and SGR models~\cite{fabry2001signal,sollich1997rheology}, and where $Y_0$ is a scaling frequency. In the SGR model, for example, $Y_0$ represents the maximum possible yielding frequency of a mesoscopic region~\cite{sollich1997rheology}. We assume that $a$ is Gaussian with average $\bar{a}$ and standard deviation $\sigma_a$ for a given study (see Fig.~\ref{fig:PLR-SDgeo3} for reported studies). We use the change-of-variables equation
$P[G(\omega)]=|da/dG(\omega)|P(a)$
to obtain
\begin{multline}P[G(\omega)]=\frac{1}{G(\omega)\sqrt{2\pi}}\frac{1}{\left|\sigma_a\left(\ln\frac{Y_0}{\omega}-\frac{d\ln f(a)}{da}\right)\right|}\times \\ \exp\left(-\frac{\left[\ln G(\omega)/f(a)-\overline{\ln G(\omega)/f(a)}\right]^2}{2\left(\sigma_a\ln \frac{Y_0}{\omega}\right)^2}\right),\end{multline} a variant of the log-normal distribution in which the place of the standard deviation is taken by the absolute value term. When $f(a)$ is only weakly dependent on $a$, as is the case for the storage modulus (Table~\ref{table:SD}), we have $d\ln f(a)/da\approx 0$ and the standard deviation is approximately  $\sigma_a\ln(Y_0/\omega)$. This standard deviation corresponds to a geometric standard deviation \SD\ $\approx(Y_0/\omega)^{\sigma_a}$ for this mechanical parameter. When this approximation is not valid (specifically, for the loss modulus), and when $d\ln f(a)/da<\ln (Y_0/\omega)$ (which applies for typical experimental condition $\omega\ll Y_0$), the base term $(Y_0/\omega)^{\sigma_a}$ is multiplied by a sub-unity correction factor $\exp[-\sigma_a \,d\ln f(a)/da]\approx\exp(\sigma_a/\bar{a})$. We list the resulting complete \SD\ terms and first-order, Taylor-series-expanded approximations for $a> 0$ in Table~\ref{table:SD}.

\begin{table*}
\centering
\caption{Predicted geometric standard deviation \SD\ of log-normal distributions for different mechanical parameters in the context of cell power-law rheology, where complex modulus increases with frequency as $\omega^a$ and creep compliance and reciprocal relaxation modulus increase with time as $t^a$, where the power-law exponent $a> 0$.}
\vspace{0.05 in}
\begin{footnotesize}
\begin{raggedright}
\begin{tabular}{lccl}
\hline \rule{0pt}{3ex}
Mechanical parameter & Form & Prefactor $f(a)$~\cite{fabry2001scaling,sollich1997rheology,fielding2000aging} & \quad Geometric standard deviation \SD\  \\[3pt] \hline \rule{0pt}{5ex}
Storage modulus $G^\prime(\omega)$ & $\propto\!f(a)\left(\dfrac{\omega}{Y_0}\right)^{a}$ & $\Gamma(1+a)\Gamma(1-a)\cos\left(\dfrac{\pi a}{2}\right)$ & $\quad\exp\left(-\sigma_a\dfrac{d\ln f(a)}{da}\right)\left(\dfrac{Y_0}{\omega}\right)^{\sigma_a}\approx \left(\dfrac{Y_0}{\omega}\right)^{\sigma_a}$ \\ [9pt]
\hline \rule{0pt}{5ex}
Loss modulus $G^{\prime\prime}(\omega)$ & $\propto\!f(a)\left(\dfrac{\omega}{Y_0}\right)^{a}$ & $\quad \Gamma(1+a)\Gamma(1-a)\sin\left(\dfrac{\pi a}{2}\right)\quad$ & $\quad\exp\left(-\sigma_a\dfrac{d\ln f(a)}{da}\right)\left(\dfrac{Y_0}{\omega}\right)^{\sigma_a}\approx \left[e^{-1/\bar{a}}\left(\dfrac{Y_0}{\omega}\right) \right]^{\sigma_a}$ \\ [9pt]
\hline \rule{0pt}{5ex} Creep compliance $J(t)$ ($t\ll t_w$)  & $\propto\!f(a)(Y_0t)^{a}$ & $\dfrac{1}{\Gamma (1+a)^2\Gamma (1-a)}$ &  $\quad\exp\left(\sigma_a\dfrac{d\ln f(a)}{da}\right)(Y_0 t)^{\sigma_a}\approx (Y_0t)^{\sigma_a}$  \\ [9pt]
\hline \rule{0pt}{5ex} Creep compliance $J(t)$ ($t\gg t_w$)  & $\propto\!f(a)(Y_0t)^{a}$ & $\dfrac{1}{\Gamma (1+a)^2\Gamma (1-a)-\Gamma (1+a)}$ &  $\quad\exp\left(\sigma_a\dfrac{d\ln f(a)}{da}\right)(Y_0 t)^{\sigma_a}\approx \left(e^{-2/\bar{a}}Y_0 t\right)^{\sigma_a}$  \\ [9pt]
\hline \rule{0pt}{5ex} $\dfrac{1}{\textrm{Relaxation~modulus~}G(t)}$ $(t\ll t_w)\quad$ & $\propto\!f(a)(Y_0t)^{a}$ & $\dfrac{1}{\Gamma (1+a)}$ &  $\quad\exp\left(\sigma_a\dfrac{d\ln f(a)}{da}\right)(Y_0 t)^{\sigma_a}\approx (Y_0t)^{\sigma_a}$  \\ [9pt]\hline
\end{tabular}
\label{table:SD}
\end{raggedright}
\end{footnotesize}
\end{table*}

We support the model's framework with experimental data obtained by optical stretching of cells in the fully suspended state; this technique involves no physical probe-cell contact, making it a favorable method for evaluating intrinsic cell-to-cell mechanical variation without conflating variation from probe contact. Adult human mesenchymal stem cells were cultured and deformed, and cell deformation analyzed, as described previously~\cite{maloney2010}, except that the cells were interrogated in the frequency domain by irradiating them with a 0.5\,Hz sinusoid with a mean power of 1\,W per fiber and a peak-to-peak value of 1\,W per fiber. Cell stiffness $G$ was calculated as $\sigma_0/\varepsilon_0$, where $\sigma_0$ is the amplitude of the photonic stress and $\varepsilon_0$ is the amplitude of a sinusoid fitted to cell deformation; the power-law exponent $a$ was calculated as $2\phi/\pi$ where $\phi$ is the phase lag of cell deformation.

As shown in Fig.~\ref{fig:G-vs-a}, our oscillatory measurements produced a trend of decreasing stiffness $G$ with increasing power-law exponent $a$. This relationship is predicted by SGR theory (because a larger agitation energy implies relatively more deformation for a given load~\cite{sollich1997rheology}). Such a prediction has not been demonstrated previously for a single experimental condition due to the conflation of probe attachment variation that is possible with other approaches; in fact, this probe engagement has been reported to introduce the opposite correlation~\cite{fabry2001signal,fabry2003time,puig-cytoskeletal}. Balland et al.'s model of log-normal distribution origin, which employs randomly selected groups of viscoelastic components, replicates this opposite correlation (i.e., increasing $G$ with increasing $a$)~\cite{balland2006power}. In contract, our focus in this work is inherent cell-to-cell mechanical variation alone; in this context, the trend we show in Fig.~\ref{fig:G-vs-a} and describe mathematically above is internally consistent and is also compatible with the SGR framework.

\begin{figure}[b!]
\includegraphics[width=8.7cm]{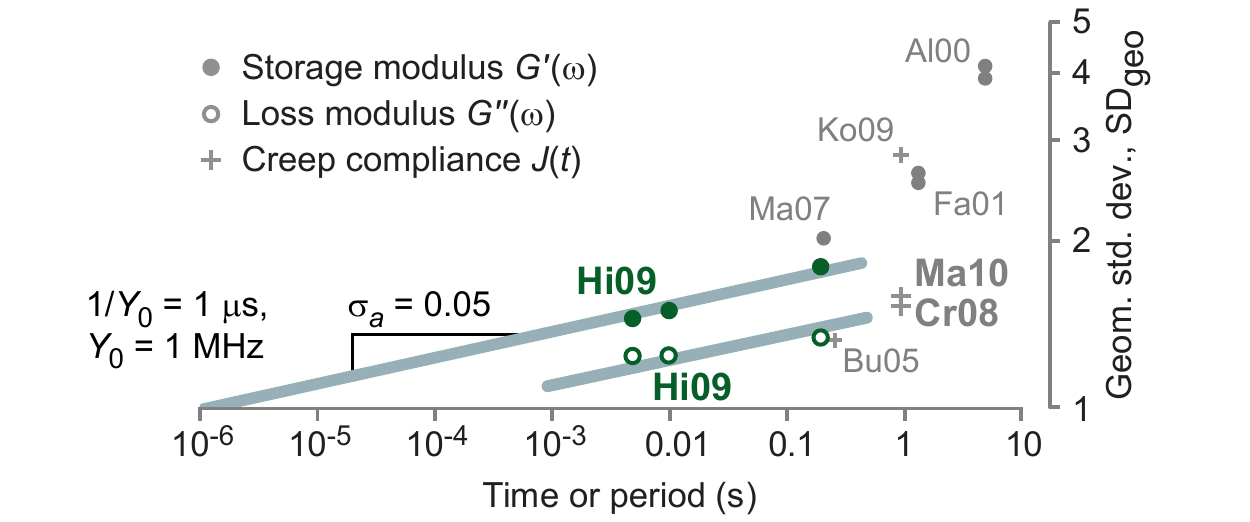}
\caption{Estimates of intrinsic variation lead to predictive models. The observed variation in cell stiffness moduli, as quantified by geometric standard deviation \SD\ of a log-normal fit, as a function of experimental time scale, for studies of \textgreater 100 cells. Solid lines show our predictions relating \SD\ to frequency for storage (upper) and loss (lower) moduli for measurements of the same cell population at different frequencies~\cite{hiratsuka2009number}, according to equations described in the text. The slopes correspond to $\sigma_a=0.05$ (best fit), in agreement with the estimate made in Fig.~\ref{fig:PLR-SDgeo3}. Data referenced by study author and year~\cite{alenghat2000analysis,fabry2001signal,
bursac2005cytoskeletal,massiera2007mechanics,cross2007nanomechanical,cross2008afm,hiratsuka2009number,kollmannsberger-nonlinear,maloney2010}; bolded references~\cite{cross2008afm,hiratsuka2009number,maloney2010} show studies  not requiring transmembrane probe-cytoskeleton linkages. (The stochastic nature of cytoskeletal engagement and anchoring tends to introduce additional experimental variation~\cite{fabry2001signal,fabry2003time,puig-cytoskeletal} not addressed by our model.)}\label{fig:PLR-SDgeo4}
\end{figure}

Having constructed a connection between the power-law-uncertainty $\sigma_a$ and the geometric standard deviation \SD\ of mechanical measurements for complex modulus $G^\star(\omega)=G^\prime(\omega)+iG^{\prime\prime}(\omega)$, we now compare our resulting predictions to experimental findings from other groups, represented in Fig.~\ref{fig:PLR-SDgeo3} and \ref{fig:PLR-SDgeo4}. Four predictions follow: First, a log-log plot of \SD\ vs.\ measurement period $1/\omega$ will have a slope $\sigma_a$ for any group of measurements.  Second, extrapolation to $\mathrm{SD}_\mathrm{geo}=1$ on this plot will occur at time 1/$Y_0$. Third, the correction factor derived above will be approximately 1 for $G^\prime$ and approximately \,$\exp(-\sigma_a/\bar{a})$ for $G^{\prime\prime}$ (at frequencies $\omega\ll Y_0$, where this restriction is explained in the derivation above). For example, we predict that the \SD\ for $G^{\prime\prime}$ will be approximately 20\% less than for $G^\prime$ for  $\bar{a}\approx 0.20$ and $\sigma_a\approx 0.05$. (The lower variation of $G^{\prime\prime}$ relative to $G^\prime$ can be understood by considering the hysteresivity $G^{\prime\prime}/G^\prime=\tan(\pi a/2)$ in the structural damping model. Larger values of $a$ reduce stiffness, but increase $G^{\prime\prime}$ relative to $G^\prime$. Therefore, fluctuations in $a$ are naturally suppressed in $G^{\prime\prime}$.) Fourth, \SD\ will generally increase with increasing period $1/\omega$; conversely, the distribution of mechanical measurements will become more narrow with increasing measurement frequency. 

All four of these predictions are confirmed by results and trends previously reported in the literature but unexplained up to this point (Fig.~\ref{fig:PLR-SDgeo3} and \ref{fig:PLR-SDgeo4}). We have extracted $\sigma_a$ and \SD\ values from multiple cell rheology reports, including a set of oscillatory indentation measurements of $G^\prime$ and $G^{\prime\prime}$ by Hiratsuka et al.~\cite{hiratsuka2009number}. Figure~\ref{fig:PLR-SDgeo4} shows a plot of \SD\ values of $G^\prime$ and $G^{\prime\prime}$ extracted from Hiratsuka et al.'s presentation of distributions at different frequencies, along with other groups' values. By fitting our model, we obtain the slope and unity intercept estimates $\sigma_a\approx 0.05$ and $1/Y_0\approx 1$\,$\upmu$s. The first fitted parameter is in good agreement with our estimate for intrinsic exponent variation described above, with $\sigma_a$  slightly less than the standard deviation (0.064, see Fig.~\ref{fig:PLR-SDgeo3}) of that group's reported $a$ values. This finding supports our first prediction above. Second, the estimate of $Y_0\approx 1$\,MHz is in fairly good agreement with Fabry et al.'s estimate~\footnote{$Y_0$ has previously been difficult to measure experimentally; Fabry et al.'s measurements on multiple cell lines range from \textless 10$^3$\,Hz to \textgreater 10$^8$\,Hz. For the cell type tested under the most conditions, human airway smooth muscle cells, the 95\% confidence interval for $Y_0$ was [1.5, 7.8]\,MHz~\cite{fabry2003time}}. Third, Hiratsuka et al.\ found \SD\ for $G^{\prime\prime}$ to be 14--25\% less than that for $G^{\prime}$(Fig.~\ref{fig:PLR-SDgeo4}), also in agreement with our  prediction of how \SD\ should vary between the real and imaginary components of the complex modulus $G^\star$. Fourth, the reported or extracted \SD\ generally does increase with period for  \SD\ values reported in the literature.

The distribution widths for creep compliance $J(t)$ and relaxation modulus $G(t)$ are readily calculated in the same way as the complex modulus. Different prefactors are possible for each of $J(t)$ and $G(t)$ depending on whether $t\ll t_w$ or $t\gg t_w$, where $t_w$ is the time between a fluidizing large-strain event and the start of the  experiment, as modeled by Fielding et al.~\cite{fielding2000aging} and experimentally explored in live-cell studies conducted by Bursac et al.\ and Trepat et al.~\cite{bursac2005cytoskeletal,trepat2007universal}. Despite these additional scenarios, the determination of \SD\  parallels our earlier treatment of the complex modulus. The resulting \SD\ terms of $G(t)$ and $J(t)$ and first-order, Taylor-series-expanded approximations for $a> 0$ are listed in Table~\ref{table:SD} as predictions to be verified~\footnote{It is assumed that $d\ln f(a)/da<\ln (Y_0 t)$, which applies for typical experimental condition $t \gg 1/Y_0$. We show only the more experimentally relevant case of $t\ll t_w$ for the relaxation modulus $G(t)$; at long times, $G(t)\rightarrow 0$.}.

Revisiting the natural questions arising from universal observations of cell populations, we find that an assumed distribution of power-law exponent $a$---together with prefactors calculated from phenomenological and theoretical models of cell rheology---lets us calculate the distribution of dynamic stiffness, creep compliance, and stress relaxation modulus. A Gaussian distribution of $a$ leads immediately to a log-normal distribution of stiffness (and of compliance, due to the reciprocal nature of this distribution). We also find the minimum measurable geometric standard deviation \SD\ for experiments conducted in the frequency regime to be $\propto$\,$(Y_0/\omega)^{\sigma_a}$. The prefactor here depends upon the experimental regime, mechanical parameter of interest, and cell state, as listed in Table~\ref{table:SD}. We have presented a new way to estimate $\sigma_a$ and $Y_0$ from the log-log slope and unity intercept of \SD\ as a function of experiment frequency. (Note that fuller interpretation of these estimates is enabled when \SD\ is reported over a range of frequencies in a given study, as in Ref. 12.) Additionally, we have predicted and confirmed an unexpected relationship between the distribution widths of $G^\prime$ and $G^{\prime\prime}$ measurements for a given cell population. Finally, given that the parameter $a$ has been linked to average activation energy or barrier height in the energy landscapes of so-called soft glassy materials, we conclude that fluctuations in $a$ (among cells and/or over time in individual cells) could plausibly produce the cell-to-cell mechanical variation that is universally observed. Our analytical model differs from previous investigations~\cite{balland2006power} that reproduce distribution shapes via phenomenological viscoelastic models, now providing  testable  predictions that relate physical mechanisms (e.g., attempt frequency) to measured distributions. 

Our framework is applicable to soft glassy regions---animate and inanimate---that exhibit PLR, as long as the standard deviation of the power-law exponent quantifies the dominant source of mechanical variation. These requirements are concluded herein to be met by living cells that are measured individually and reported as a population distribution that is consequently log-normal. Our model focuses on endogenous variation and its manifestation as measured by low-contact or no-contact techniques (such as oscillatory probe indentation~\cite{hiratsuka2009number} and optical stretching~\cite{maloney2010}, respectively), in which probe-cytoskeleton linkages are not required, thus reducing or eliminating measurement variation caused by ligand-receptor engagement  of the cytoskeleton.  Additional experiments and models are needed to determine conclusively whether disease state or chemomechanical modulation of the cell environment, both of which are known to alter cell stiffness, can also modulate the mechanical \emph{variation} among cells \footnote{For example, Cross et al.\ initially reported a normal distribution of stiffness values for cancer cells~\cite{cross2007nanomechanical}, but their follow-up study on the same primary tumor cells, prepared via a different protocol, indicates a log-normal distribution~\cite{cross2008afm}.}. Most valuable are high-throughput cell measurement techniques that allow robust estimates of distribution width as a function of time or oscillation frequency, as well as estimated errors in these widths. These quantities will inform models such as ours that can then connect microscopic structural and energetic barriers within cells to emergent changes in mechanical distributions among cells.

We gratefully acknowledge the Beckman Foundation, US NIH NIBIB training grant No.\ EB006348 (JMM), and US NSF CAREER Award No.\ CBET-0644846 (KJVV).

\end{document}